\documentclass[prb,preprint,amsmath,amssymb,superscriptaddress,notitlepage]{revtex4-1}
\usepackage{epsfig}
\usepackage{graphicx}
\usepackage{color}
\usepackage{bm}
\usepackage{fixme}
\usepackage[colorinlistoftodos,textwidth=2.25cm]{todonotes} 
\usepackage{color,soul}
\usepackage{xcolor}
\usepackage{comment}

\usepackage{tikz}
\usetikzlibrary{positioning}

\tikzstyle{box} = [rectangle, rounded corners, minimum width=3cm, minimum height=1cm,text centered, draw=black, text width=6.5cm]
\tikzstyle{arrow} = [very thick,->,>=stealth]

\newcommand\colW{0.5}

\begin{document}

\title{Compressed AFM-IR hyperspectral nanoimaging of single \emph{Leishmania} parasites.}

\author{A.~Hornemann} 
\author{M.~Marschall} 
\author{S.~Metzner} 
\affiliation{Physikalisch-Technische Bundesanstalt (PTB), Abbestr. 2-12, 10587 Berlin, Germany}
\author{P.~Patoka} 
\affiliation{Freie Universität Berlin, Arnimallee 22, 14195 Berlin, Germany}
\author{S.~Cortes} 
\affiliation{Global Health and Tropical Medicine (GHTM), Instituto de Higiene e Medicina Tropical (IHMT), Universidade Nova de Lisboa, Rua Junqueira 100, 1349-008, Lisbon, Portugal}
\author{G.~Wübbeler} 
\author{A.~Hoehl} 
\affiliation{Physikalisch-Technische Bundesanstalt (PTB), Abbestr. 2-12, 10587 Berlin, Germany}
\author{E.~Rühl} 
\email{ruehl@zedat.fu-berlin.de}
\affiliation{Freie Universität Berlin, Arnimallee 22, 14195 Berlin, Germany}
\author{B.~K\"astner} 
\email{bernd.kaestner@ptb.de}
\author{C.~Elster}
\affiliation{Physikalisch-Technische Bundesanstalt (PTB), Abbestr. 2-12, 10587 Berlin, Germany}
\maketitle

\section{Abstract}
Infrared hyperspectral imaging is a powerful approach in the field of materials and life sciences. However, the extension to modern sub-diffraction nanoimaging still remains a highly inefficient technique, as it acquires data via inherent sequential schemes. Here, we introduce the mathematical technique of low-rank matrix reconstruction to the sub-diffraction scheme of atomic force microscopy-based infrared spectroscopy (AFM-IR), for  efficient hyperspectral infrared nanoimaging.
To demonstrate its application potential, we chose the trypanosomatid unicellular parasites \emph{Leishmania} species as a realistic target of biological importance. The mid-infrared spectral fingerprint window covering the spectral range from 1300 to 1900 cm$^{-1}$ was chosen and a step width of 220\,nm was applied for nanoimaging of single parasites. Multivariate statistics approaches such as hierarchical cluster analysis (HCA) were used for extracting the chemically distinct spatial locations. 
Subsequently, we randomly selected only 5\,\% from an originally gathered data cube of 134 (x) $\times$ 50 (y) $\times$ 148 (spectral) AFM-IR measurements and reconstructed the full data set by low-rank matrix recovery.
The technique is evaluated by showing agreement in the cluster regions between full and reconstructed data cubes. We conclude that the corresponding measurement time of more than 14 hours can be reduced to less than 1 hour.
These findings can significantly boost the practical applicability of hyperspectral nanoimaging in both academic and industrial settings involving nano- and bio-materials.

\section{Introduction}

Infrared (IR) hyperspectral imaging (HSI), i.e. the recording of a spectrum at each pixel of a 2D specimen, is a powerful approach for non-invasive and non-ionizing materials characterization ranging from analytical chemistry~\cite{salzer2009,morsch2020}, materials sciences \cite{ruggeri2015,chae2015}, life sciences \cite{pilling2016,PALUSZKIEWICZ2017}  to microelectronics~\cite{lau1999}. 
HSI allows for directly correlating sample's morphology and topography with spectral signal response and offers the great capability of high-throughput spatially resolved analysis when combined with multivariate statistics and machine learning tools, thus, enabling automated detection, pattern recognition, and phenotyping, particularly in the field of biomedical diagnostics \cite{lu2014, Lasch2018, hu2019}.

Hyperspectral maps are usually obtained from parallel detection, for instance by employing  focal-plane-array (FPA) detectors, which can considerably reduce data acquisition times via simultaneous detection of multi-pixel elements \cite{levin2005}.  However, the highest spatial resolution of conventional optical techniques is limited by diffraction \cite{Chan2018}. The diffraction limit can be overcome by modern scanning-based methods, such as atomic force microscopy-based IR spectroscopy (AFM-IR)~\cite{dazzi2005, Dazzi2017}, scattering-type scanning near-field optical microscopy (s-SNOM)~\cite{knoll1999, keilmann2009}, photoinduced force microscopy (PiFM)~\cite{Rajapaska2010}, and tip-enhanced Raman spectroscopy (TERS)~\cite{STOCKLE2000131}. As these approaches gather spectra in a sequential manner, HSI is highly challenging for a larger 2D array. 

Typically, HSI measurements using scanning based methods may take several hours
to reach a meaningful detection sensitivity (see, e.g., Ref.~\onlinecite{amenabar2017}). This, on the other hand, may lead to sample and tip damage as well as drift artifacts \cite{Canale2011}. This compromise between sensitivity and data acquisition time inhibits unfolding the potential of HSI for reliably identifying and distinguishing chemical species.
To overcome this limitation, we introduce the mathematical technique of low-rank matrix reconstruction to the sub-diffraction scheme of atomic force microscopy-based infrared spectroscopy (AFM-IR), for efficient hyperspectral infrared nanoimaging. 

\newpage
\section{Compressed AFM-IR spectral imaging}
The AFM-IR based scanning technique allows for a label-free analysis for mapping both the morphology and spatially resolved fingerprint signatures of absorbing molecular species in biological matrices, resulting in high-resolution chemical images \cite{quaroni2018,dazzi2012}.
AFM-IR is considered to be a true model-free IR absorption spectroscopy with a penetration depth of the IR beam in the order of several micrometers, providing spectral signatures that resemble those in the far-field-FTIR regime.

First attempts in emulating a measurement setting with significantly reduced acquisition time have been made. This was done by taking recourse of interferograms originating from hyperspectral data sets of scanning methods such as Fourier-transform infrared spectroscopy (FTIR),
demonstrating that it is possible to significantly reduce the measurement time by applying the mathematical technique of low-rank matrix reconstruction \cite{Wuebbeler2021, marschall2020}.
This compression of such signal sources is possible because of their redundancy without substantially losing information. 

For introducing the method of low-rank matrix reconstruction~\cite{Candes2008, Davenport2016, marschall2020} into the field of AFM-IR spectroscopy we apply it to hyperspectral infrared (IR) nanoimaging of single \emph{Leishmania} parasites as a realistic target of biological importance \cite{WHO2022}. 
\textit{Leishmania} responsible for the vector-borne disease leishmaniasis present two morphological forms along its life cycle in the vertebrate host (amastigotes) and insect vector (promastigotes). We conducted hyperspectral nanoimaging of promastigote forms of \textit{L. brasiliensis} which are spindle-shaped with a \textit{flagellum}. The low-rank approximation concept is applied to original unprocessed AFM-IR hyperspectral data gathered from a photothermal expansion-based AFM-IR setup. We demonstrate a thorough evaluation of the low-rank matrix recovery method
by a reduction to only  5$\%$ of randomly  chosen measurements from the complete (original) data set. Furthermore, a subsequent data analysis and biochemical interpretation was carried out
multivariately (hierarchical cluster analysis imaging) by comparing the information content for the reconstructed and full (original) hyperspectral data set.

\vspace*{1cm}

\begin{figure}[t]
\hspace*{-0.39cm}
\includegraphics[width=1.05\textwidth]{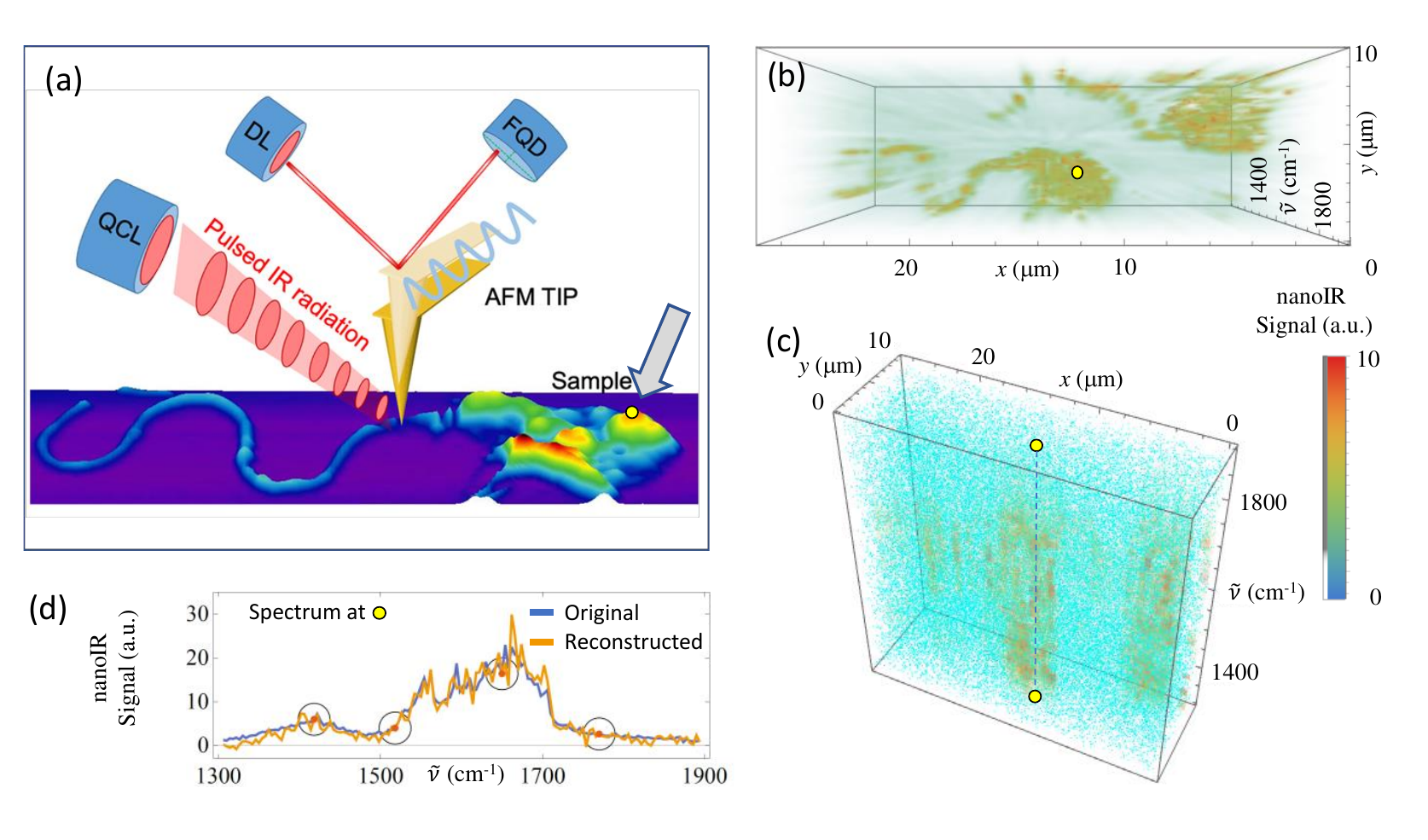}
\caption
{{\bf (a)} Photothermal expansion experimental setup: QCL – tunable quantum cascade laser, DL – deflection laser, FQD – four-quadrant photodiode. The topography of the sample is shown, with the arrow pointing to the indicated sampe spot marked by a yellow circle. {\bf (b)} 3D representation of the data set viewed along the spectral axes. The color scale shown below encodes the signal strength both as color and transparency, so that the spatial distribution of two \emph{Leishmania} promastigotes can be seen. The yellow circle corresponds to the indicated measurement position in (a). {\bf (c)} Illustration of the compressed measurement, showing the randomly chosen voxels in the data  cube, to which low-rank matrix recovery was applied. The dashed line connecting the two yellow points corresponds to the infrared spectrum shown in (d). {\bf (d)} nanoIR spectrum at the position marked by a yellow circle. The original and the reconstructed data set are indicated by blue and orange curves, respectively. The marked points belong to the randomly chosen set of data used for the reconstruction.}
\label{fig-Main}
\end{figure}

The spectral measurements of \emph{Leishmania brasiliensis} promastigotes were carried out with a nanoIR2-s (Bruker (former Anasys), USA) photothermal expansion-based microscope combining atomic force microscopy with a local thermal expansion effect\cite{mathurin2022, Rajes2021}. A schematic of the setup is shown in Fig.\,\ref{fig-Main}(a). During the experiments pulsed IR laser light from a tunable quantum cascade laser (QCL)(MIRcat-QT\textsuperscript{TM}, DRS Daylight Solutions Inc., USA) was focused onto the sample at the proximity of the AFM tip. The corresponding absorption results in local rapid thermal expansion under the condition that the wavelength of the IR laser radiation matches the absorption bands of the samples under study. The induced expansion of the sample generates oscillations of the cantilever which are registered by changes of the deflection laser (DL) reflection spot on the four-quadrant photodiode (FQD) of the AFM. High sensitivity was achieved by using the resonance enhanced mode. In this mode the repetition rate of the laser is continuously matched to the contact resonance frequency of the cantilever in contact with the sample. Local absorption spectroscopy is realized by recording changes of the cantilever oscillation amplitude as a function of infrared laser emission wavelength. In HSI the full spectral information is gathered in each pixel of the topographic map, allowing for a local chemical analysis of the sample related to its topography.

The principle of a compressed measurement using AFM-IR imaging is shown in Fig.\,\ref{fig-Main}(b)-(d). AFM-IR spectral fingerpints were collected via HSI data acquisition in the 1900 to 1300~cm$^{-1}$ spectral window from two single \emph{L. brasiliensis} parasites (promastigote forms). An illustation of the full HSI data set as a 3D data cube is shown in Fig. 1(b). It consists of so called voxels formed by 134 x 50 points along the $(x, y)$ axes as spatial and 148 points along the $\tilde{\nu}$ axis as the spectral coordinate. The spectral resolution is approximately 4\,cm$^{-1}$, with a data acquisition time of 8 seconds per spectrum. The total time for recording the full data cube  was approximately 14 h.

The projection shown in Fig.\,\ref{fig-Main}(b) views the data cube along the spectral axis, so that the two spectrally integrated promastigotes can be seen. As an example, the yellow circle in Fig.\,\ref{fig-Main} indicates an arbitrarily chosen position to illustrate the data evaluation. 

In a compressed measurement only a small fraction of randomly selected voxels of the data cube are used. This is shown in the blue dots in Fig.\,\ref{fig-Main}(c). In this example 5\% of the full data set are used. In particular, this means that only a fraction of wavenumbers are considered, as illustrated by the dashed line connecting the upper and lower yellow circle. Using only 5\% of the data means that, in principle, the data acquisition  time can be reduced to less than 1 hour. Translation of data reduction into a similar reduction of measurement time is possible by continuously measuring and moving all three axes which will be discussed elsewhere~\cite{Metzner2022}.

By applying the method of low-rank matrix reconstruction~\cite{marschall2020, kaestner2018,Davenport2016}, as detailed below, the full spectrum can be retrieved as long as the measurement fulfills certain redundancy properties which is often the case in HSI measurements \cite{chen2020}. Fig.\,\ref{fig-Main}(d) shows, for example, an infrared spectrum taken at the point marked by the arrow in Fig. 1(a) and yellow discs in Fig. 1(b) and Fig. 1(c). The full  and reconstructed spectra are shown in blue and orange color, respectively. Both curves are in agreement despite the fact that the reconstructed data is based on only 5$\%$ of the original data. These randomly selected points are marked by gray circles.

\section{Low-rank matrix recovery}
\label{sec:LR}

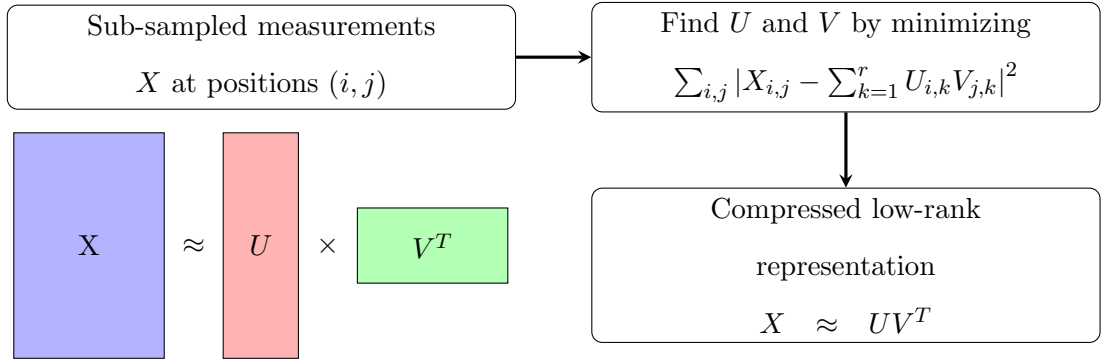
\begin{figure}
	\centering
	\begin{tikzpicture}
   \node (x) [box] {Sub-sampled measurements \\ $X$ at positions $(i, j)$};
   \node (min) [box, right=of x] {Find $U$ and $V$ by minimizing \\ $\sum_{i, j} \left\vert X_{i,j} - \sum_{k=1}^r U_{i,k} V_{j,k}\right\vert^2$};
   \node (lr) [box, below=of min] {Compressed low-rank \\ representation \\ $X\approx UV^T$};
   \draw[arrow] (x) -- (min);
   \draw[arrow] (min) -- (lr);
   \node (xmat) [rectangle, below left = 0.3cm and -2.1cm of x, minimum width=2cm, minimum height=3cm, text centered, draw=black, fill=blue!30] {X};
   \node (approx) [right= 0.1cm of xmat] {$\approx$};
   \node (umat) [rectangle, right = 0.1cm of approx, minimum width=1cm, minimum height=3cm, text centered, draw=black, fill=red!30] {$U$};
   \node (times) [right=0.1cm of umat] {$\times$};
   \node (vmat) [rectangle, right = 0.1cm of times, minimum width=2cm, minimum height=1cm, text centered, draw=black, fill=green!30] {$V^T$};
\end{tikzpicture}
	\caption{Concept of low-rank matrix recovery.}
	\label{Low-rank-scheme}
\end{figure}

Low-rank matrices arise in many settings related to mathematical modeling and data compression. Applications range from signal processing \cite{weng2012low} to image restoration \cite{peng2014reweighted} and machine learning \cite{yao2015fast}.

A relevant example of problems includes the recovery of a data matrix given only by incomplete observations~\cite{marschall2020}. The data matrix is then approximated by a matrix product with each factor having lower dimensionality, resulting in a recovery result of lower rank (c.f. Fig. ~\ref{Low-rank-scheme}). 

The main idea is that a low-rank approximation already captures the main characteristics of the data and less informative dimensions will be removed.
In Ref.~\onlinecite{marschall2020}, an algorithm for low rank matrix reconstruction has been presented and successfully applied to sub-sampled (FPA)-FTIR data. For observations $X$, the task is to find matrices $U$ and $V$ by minimizing $\sum_{i,j}|X_{i,j}-\sum_{k=1}^rU_{i,k}V_{j,k}|^2$. The rank $r$ approximation $\hat{X}$ of $X$ is then given by $\hat{X}=UV^T$. 
The problem depicted above is in general ill-posed. Hence, an additional Tikhonov regularization is utilized \cite{engl1996regularization} together with a smoothness constraint to the spatial domain. To solve the resulting minimization problem, an alternating algorithm  is employed which generates linear problems in each iteration. A Python implementation of the applied algorithm is publicly available \cite{Software1}.


\vspace*{1cm}

\section{Hyperspectral imaging of \emph{Leishmania} parasites}

In this Section, we use the example of HSI of the unicellular parasite \emph{L. brasiliensis}, responsible for cutaneous and mucocutaneous leishmaniasis \cite{WHO2022} for a spectroscopic interpretation based on both, the full and the reconstructed data set.

AFM-IR hyperspectral nanoimaging is a powerful tool for probing the composition of entire parasitic organisms, covering a dimension of about 10–15 $\mu$m, and to enable a direct correlation and/or differentiation to morphological settings, such as the anterior and posterior parts (\emph{flagellum}) of promastigote forms.

\subsection{Spectral interpretation}

The following regions were put into our focus: The 1900–1500 cm$^{-1}$ spectral window, where the following molecular constituents, such as proteins, deliver amide I and amide II bands. 
Amide bonds are abundant in proteins because of their higher stability and proclivity for forming resonating structures, which influence secondary structure adoption and biological activity \cite{mahesh2018}. Particularly, band positions, band widths (full-width-at-half-maximum, FWHM) and band shifts of amide bonds may represent an important indicator for molecular structural (re-)organizations. In contrast to NMR spectra, which provide information on the tertiary structure of proteins, IR spectra focus on providing information about the secondary structure of proteins \cite{yang2015}.
The amide window is followed by the so-called 'mixed region' (1500–1300 cm$^{-1}$), including fatty acid bending vibrations, C–N stretching and N–H deformational modes of proteins, and P=O stretching modes of phosphate-carrying species. The spectral modes are summarized in Table~\ref{tab-modes}.

\begin{table}
\begin{tabular}{cccc}
\hline
Observed modes / cm$^{-1}$ & Molecular vibration & Band assignments\\ 
\hline
 1753–1735 & $\nu$(C=O), $\nu$(COOH) & Saturated esters\\ 
 1725–1705 & $\nu$(C=O) & Ketones,–COOH\\
 1677–1640 & $\nu$(C=O) or $\nu$(C=C), $\nu$(C=N) & Amide I\\
 1640–1614 & $\delta$(N–H) & Primary amide\\ 
 1562-1550 & $\delta$(N–H), $\nu$(C–N), $\nu$(C=N) & NHR, secondary amine,\\ & & protein and nucleic acid\\
1420–1400 & $\nu$(C–N) & Primary amide\\
\hline
\end{tabular}
\caption{Overview of relevant vibrational modes detected in the MIR spectral regime \cite{parker1975,helm1991}}.
\label{tab-modes}
\end{table}

\newpage

\subsection{Hierarchical cluster analysis imaging}

Hierarchical cluster analysis (HCA) implementing Ward's method \cite{ward1963} was performed on the HSI data cubes for extracting the chemically distinct spatial assignments (see Appendix). Cluster analysis is a useful approach for multivariate differentiation and classification studies of biological specimens, and can be easily applied to IR-based hyperspectral data \cite{hornemann2017}.
Further, the spectral characteristics of the measured arithmetic mean spectrum of the respective cluster, and of the low-rank reconstructed arithmetic mean spectrum, representing each cluster, was evaluated, by applying fitting procedures to Voigt profiles, specifically in the amide regime.


\begin{figure}[t]
\hspace*{-0cm}\includegraphics[width=.9\textwidth]{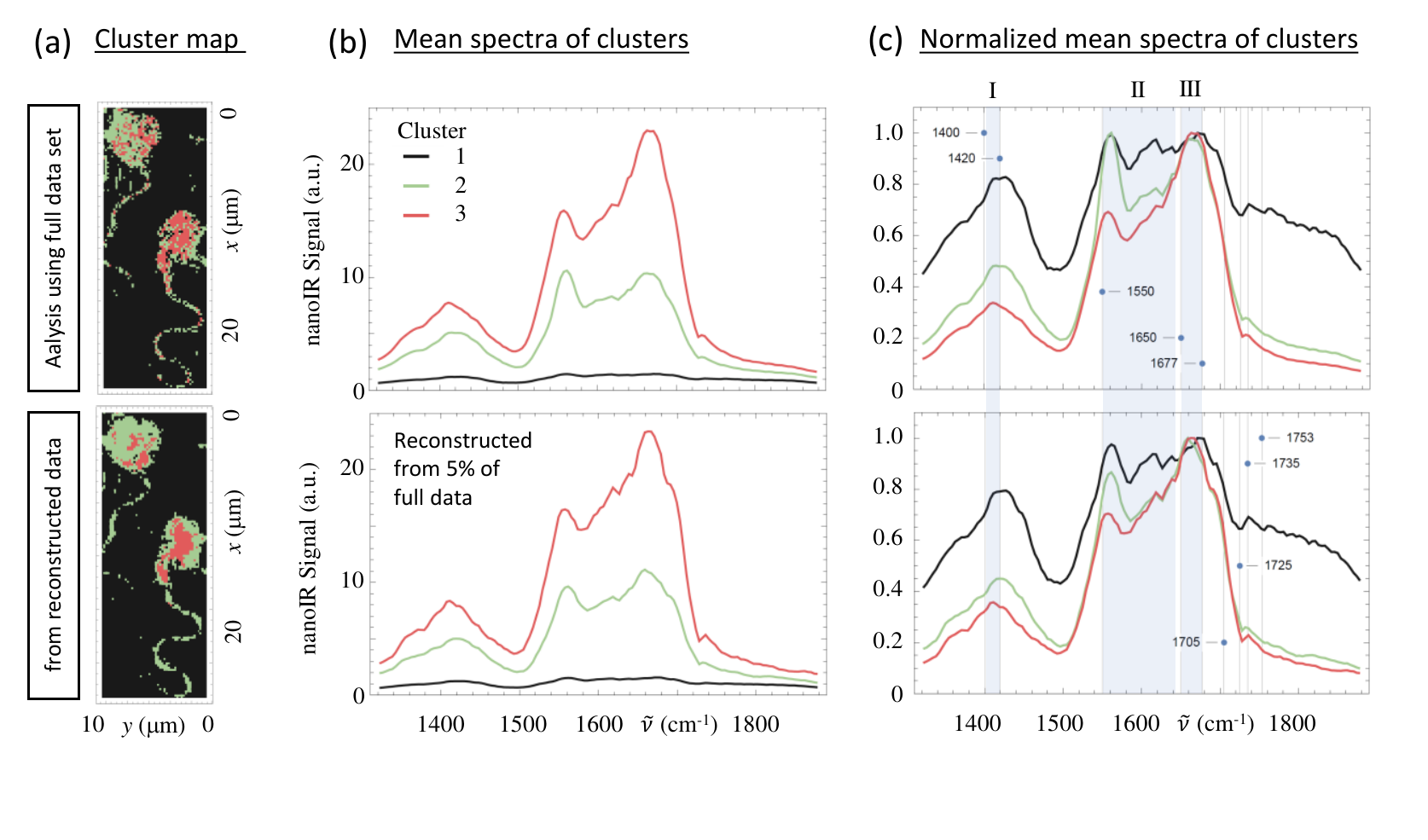}
\caption
{
Results of cluster analysis performed using the full data set (top) and the data set reconstructed from 5$\%$ of the full data set (bottom). {\bf (a)} Spatial distribution of the three spectral clusters as determined by Ward's method. {\bf (b)} Spectra averaged over each of the three clusters. {\bf (c)} Spectra as in (b) but normalized to their maximum value. The relevant spectral regions are shaded and labelled as (I): primary amides, (II) the amide II, and (III) the amide I region.
}
\label{fig-Cluster}
\end{figure}

Fig.~\ref{fig-Cluster} (a) shows the HCA-based cluster maps calculated from the original (top) and reconstructed datacubes (bottom), respectively. Both HCA maps of the two \textit{Leishmania} parasites exhibit three different clustered groups, given by the spectral background (black), spectral features located mainly in the body and the \emph{flagellum} regions (red), and the third cluster group (green), the latter represents spectral features, mainly originating from the \textit{flagellum} and marginal body regions. 
Exceptionally, the parasites' morphologies are found correctly represented in the reconstructed cluster maps, as can be seen from the \emph{flagellum} outline and body dimensions. This is a quite important aspect in view of the characterization of a biological specimen, particularly in single cell analysis, and after microscope examination for microorganisms identification, to be combined with molecular nanoscopy for a subsequent biochemical analysis.
If we compare the HCA maps obtained from the full and the reconstructed data cubes, the cluster allocations are, to a large extent, recovered at nearly the same pixel coordinates. However, small deviations can be observed. These deviations result mainly from a smoothing effect resulting from the application of the Tikhonov regularization as mentioned in Section~\ref{sec:LR}. 

A comparison between full (top) and reconstructed (bottom) arithmetic mean spectra (cf. Fig.~\ref{fig-Cluster} (b)) indicates well reproduced spectral profiles, considered for each cluster group. The band shapes and ratios, as well as their FWHM, could be successfully reproduced on basis of a low rank reconstruction using only 5$\%$ of the data. Furthermore, the baselines, and spectral offsets, can be completely recovered from low-rank modeling with respect to their magnitudes. 

\newpage

Fig.~\ref{fig-Cluster} (c) displays a comparison of full spectra and reconstructed signatures for the corresponding cluster groups, respectively, which were normalized to their maximum. This was done in order to determine the peak ratios between full and reconstructed arithmetic mean spectra. The following regions of interest, highlighted here in gray color, are discussed in the following: the region of primary amides referring to C–N stretching vibrations located at 1420 cm$^{-1}$–1400 cm$^{-1}$ (I), the amide II region at 1550–1640 cm$^{-1}$ (II), and the amide I region found at 1650–1677cm$^{-1}$ (III).
Basically, cluster arithmetic mean spectra of original data cubes resemble the reconstructed arithmetic mean profiles, and particularly the peak ratios are in line with each other, i.e. the mean spectrum of cluster 3 (red lines), which features an enhancement of the amide I/amide II band ratios compared to the remaining cluster profiles (black and green lines), can be correctly modeled. 

Also, the slight spectral shift of the amid II region between cluster 2 and 3 was reproduced. To show this we performed nonlinear curve fitting to Voigt Profiles (see Appendix). The band position of cluster 2 and 3 determined from the original (reconstructed) data was 1561 (1562) cm$^{-1}$ and 1559 (1559) cm$^{-1}$, respectively. Note also, that the spectral features around 1753-1735 cm$^{-1}$ referring to the baselines turn up again at the same position.
The results have proven that the low-rank methodology provides a reliable modeling tool to 
reconstruct multi-dimensional data of complex biomatrices, as not only the protein content lying behind the amide I/amide II bands, and their ratios, could be entirely recovered, but also their spectral background and offsets.
The latter is likely due to wavenumber-dependent scattering or non-compensated reflection losses \cite{centrone2015}.
Furthermore, the signal-to-noise ratios of the reconstructed data could be successfully reproduced.

\section{Conclusions}

We presented for the first time a procedure for compressed AFM-IR spectroscopy using low-rank reconstruction for efficient hyperspectral infrared nanoimaging. The suitability and application potential was demonstrated using signatures of single parasitic specimens \textit{L. brasiliensis}. We have shown that the use of only 5$\%$ of randomly selected data from the original data cube enabled an adequate reconstruction of the entire data cube. The MIR fingerprint spectral complexity could be successfully reproduced, specifically in the amide and "mixed" regions. This was evaluated by comparing HCA maps leading to the same bio-analytical information.
We conclude that the data acquisition times required for full hyperspectral measurements in AFM-IR procedures may be significantly reduced by low-rank matrix reconstruction schemes, promoting scanning based methods for hyperspectral imaging which so far only parallel schemes qualify for.  

\begin{acknowledgements}
We gratefully acknowledge fruitful discussions with Alexander Govyadinov. 
We acknowledge financial support by Deutsche Forschungsgemeinschaft (Grants EL 492/1-1, RU 420/13-1).
Sofia Cortes was supported by Fundação para a Ciência e Tecnologia, through project UID/Multi/04413/2019 (GHTM).
\end{acknowledgements}

\newpage
\bibliographystyle{naturemag}
\bibliography{Refs.bib}

\begin{thebibliography}{10}
\expandafter\ifx\csname url\endcsname\relax
  \def\url#1{\texttt{#1}}\fi
\expandafter\ifx\csname urlprefix\endcsname\relax\def\urlprefix{URL }\fi
\providecommand{\bibinfo}[2]{#2}
\providecommand{\eprint}[2][]{\url{#2}}

\bibitem{salzer2009}
\bibinfo{author}{Salzer, R.} \& \bibinfo{author}{Siesler, H.~W.}
\newblock \emph{\bibinfo{title}{{Infrared and Raman Spectroscopic Imaging}}}
  (\bibinfo{publisher}{Wiley}, \bibinfo{year}{2009}).

\bibitem{morsch2020}
\bibinfo{author}{Morsch, S.}, \bibinfo{author}{Lyon, S.},
  \bibinfo{author}{Edmondson, S.} \& \bibinfo{author}{Gibbon, S.}
\newblock \bibinfo{title}{{Reflectance in AFM-IR: Implications for
  Interpretation and Remote Analysis of the Buried Interface}}.
\newblock \emph{\bibinfo{journal}{Anal. Chem.}} \textbf{\bibinfo{volume}{92}},
  \bibinfo{pages}{8117--8124} (\bibinfo{year}{2020}).

\bibitem{ruggeri2015}
\bibinfo{author}{Ruggeri, F.~S.} \emph{et~al.}
\newblock \bibinfo{title}{Infrared nanospectroscopy characterization of
  oligomeric and fibrillar aggregates during amyloid formation}.
\newblock \emph{\bibinfo{journal}{Nature Communications}}
  \textbf{\bibinfo{volume}{6}}, \bibinfo{pages}{7831}.

\bibitem{chae2015}
\bibinfo{author}{Chae, J.}, \bibinfo{author}{Dong, Q.}, \bibinfo{author}{Huang,
  J.} \& \bibinfo{author}{Centrone, A.}
\newblock \bibinfo{title}{{Chloride Incorporation Process in
  CH$_3$NH$_3$PbI$_{3-x}$Cl$_x$ Perovskites via Nanoscale Bandgap Maps}}.
\newblock \emph{\bibinfo{journal}{Nano Lett.}} \textbf{\bibinfo{volume}{15}},
  \bibinfo{pages}{8114--8121} (\bibinfo{year}{2015}).

\bibitem{pilling2016}
\bibinfo{author}{Pilling, M.} \& \bibinfo{author}{Gardner, P.}
\newblock \bibinfo{title}{Fundamental developments in infrared spectroscopic
  imaging for biomedical applications}.
\newblock \emph{\bibinfo{journal}{Chem. Soc. Rev.}}
  \textbf{\bibinfo{volume}{45}}, \bibinfo{pages}{1935--1957}
  (\bibinfo{year}{2016}).

\bibitem{PALUSZKIEWICZ2017}
\bibinfo{author}{Paluszkiewicz, C.} \emph{et~al.}
\newblock \bibinfo{title}{{Differentiation of protein secondary structure in
  clear and opaque human lenses: AFM – IR studies}}.
\newblock \emph{\bibinfo{journal}{Journal of Pharmaceutical and Biomedical
  Analysis}} \textbf{\bibinfo{volume}{139}}, \bibinfo{pages}{125--132}
  (\bibinfo{year}{2017}).

\bibitem{lau1999}
\bibinfo{author}{Lau, W.~S.}
\newblock \emph{\bibinfo{title}{{Infrared Characterization for
  Microelectronics}}} (\bibinfo{publisher}{World Scientific Publishing},
  \bibinfo{year}{1999}).

\bibitem{lu2014}
\bibinfo{author}{Lu, G.} \& \bibinfo{author}{Fei, B.}
\newblock \bibinfo{title}{Medical hyperspectral imaging: a review.}
\newblock \emph{\bibinfo{journal}{J Biomed Opt}} \textbf{\bibinfo{volume}{19}},
  \bibinfo{pages}{10901} (\bibinfo{year}{2014}).

\bibitem{Lasch2018}
\bibinfo{author}{Lasch, P.} \emph{et~al.}
\newblock \bibinfo{title}{{FT-IR Hyperspectral Imaging and Artificial Neural
  Network Analysis for Identification of Pathogenic Bacteria.}}
\newblock \emph{\bibinfo{journal}{Analytical chemistry}}
  \textbf{\bibinfo{volume}{90}}, \bibinfo{pages}{8896--8904}
  (\bibinfo{year}{2018}).

\bibitem{hu2019}
\bibinfo{author}{Hu, B.}, \bibinfo{author}{Du, J.}, \bibinfo{author}{Zhang, Z.}
  \& \bibinfo{author}{Wang, Q.}
\newblock \bibinfo{title}{Tumor tissue classification based on
  micro-hyperspectral technology and deep learning}.
\newblock \emph{\bibinfo{journal}{Biomed. Opt. Express}}
  \textbf{\bibinfo{volume}{10}}, \bibinfo{pages}{6370--6389}
  (\bibinfo{year}{2019}).

\bibitem{levin2005}
\bibinfo{author}{Levin, I.~W.} \& \bibinfo{author}{Bhargava, R.}
\newblock \bibinfo{title}{Fourier transform infrared vibrational spectroscopic
  imaging: Integrating microscopy and molecular recognition}.
\newblock \emph{\bibinfo{journal}{Annual Review of Physical Chemistry}}
  \textbf{\bibinfo{volume}{56}}, \bibinfo{pages}{429--474}
  (\bibinfo{year}{2005}).

\bibitem{Chan2018}
\bibinfo{author}{Chan, K. L.~A.}, \bibinfo{author}{Fale, P. L.~V.},
  \bibinfo{author}{Atharawi, A.}, \bibinfo{author}{Wehbe, K.} \&
  \bibinfo{author}{Cinque, G.}
\newblock \bibinfo{title}{Subcellular mapping of living cells via synchrotron
  microftir and zns hemispheres}.
\newblock \emph{\bibinfo{journal}{Analytical and Bioanalytical Chemistry}}
  \textbf{\bibinfo{volume}{410}}, \bibinfo{pages}{6477--6487}
  (\bibinfo{year}{2018}).

\bibitem{dazzi2005}
\bibinfo{author}{Dazzi, A.}, \bibinfo{author}{Prazeres, R.},
  \bibinfo{author}{Glotin, F.} \& \bibinfo{author}{Ortega, J.~M.}
\newblock \bibinfo{title}{{Local infrared microspectroscopy with subwavelength
  spatial resolution with an atomic force microscope tip used as a photothermal
  sensor}}.
\newblock \emph{\bibinfo{journal}{Optics Letters}}
  \textbf{\bibinfo{volume}{30}}, \bibinfo{pages}{2388} (\bibinfo{year}{2005}).

\bibitem{Dazzi2017}
\bibinfo{author}{Dazzi, A.} \& \bibinfo{author}{Prater, C.~B.}
\newblock \bibinfo{title}{{AFM-IR: Technology and Applications in Nanoscale
  Infrared Spectroscopy and Chemical Imaging}}.
\newblock \emph{\bibinfo{journal}{Chemical Reviews}}
  \textbf{\bibinfo{volume}{117}}, \bibinfo{pages}{5146--5173}
  (\bibinfo{year}{2017}).

\bibitem{knoll1999}
\bibinfo{author}{Knoll, B.} \& \bibinfo{author}{Keilmann, F.}
\newblock \bibinfo{title}{{Near-field probing of vibrational absorption for
  chemical microscopy}}.
\newblock \emph{\bibinfo{journal}{Nature}} \textbf{\bibinfo{volume}{399}},
  \bibinfo{pages}{134} (\bibinfo{year}{1999}).

\bibitem{keilmann2009}
\bibinfo{author}{Keilmann, F.} \& \bibinfo{author}{Hillenbrand, R.}
\newblock \bibinfo{title}{{Near-field nanoscopy by elastic light scattering
  from a tip}}.
\newblock In \bibinfo{editor}{Zayats, A.} \& \bibinfo{editor}{Richard, D.}
  (eds.) \emph{\bibinfo{booktitle}{Nano-Optics and Near-Field Optical
  Microscopy}}, chap.~\bibinfo{chapter}{11}, \bibinfo{pages}{235--265}
  (\bibinfo{publisher}{Artech House, Boston, London}, \bibinfo{year}{2009}).

\bibitem{Rajapaska2010}
\bibinfo{author}{Rajapaksa, I.}, \bibinfo{author}{Uenal, K.} \&
  \bibinfo{author}{Wickramasinghe, H.~K.}
\newblock \bibinfo{title}{{Image force microscopy of molecular resonance: A
  microscope principle}}.
\newblock \emph{\bibinfo{journal}{Applied Physics Letters}}
  \textbf{\bibinfo{volume}{97}}, \bibinfo{pages}{073121}
  (\bibinfo{year}{2010}).

\bibitem{STOCKLE2000131}
\bibinfo{author}{St{\"{o}}ckle, R.~M.}, \bibinfo{author}{Suh, Y.~D.},
  \bibinfo{author}{Deckert, V.} \& \bibinfo{author}{Zenobi, R.}
\newblock \bibinfo{title}{{Nanoscale chemical analysis by tip-enhanced Raman
  spectroscopy}}.
\newblock \emph{\bibinfo{journal}{Chemical Physics Letters}}
  \textbf{\bibinfo{volume}{318}}, \bibinfo{pages}{131--136}
  (\bibinfo{year}{2000}).

\bibitem{amenabar2017}
\bibinfo{author}{Amenabar, I.} \emph{et~al.}
\newblock \bibinfo{title}{Hyperspectral infrared nanoimaging of organic samples
  based on {Fourier} transform infrared nanospectroscopy}.
\newblock \emph{\bibinfo{journal}{Nature Communications}}
  \textbf{\bibinfo{volume}{8}}, \bibinfo{pages}{14402} (\bibinfo{year}{2017}).

\bibitem{Canale2011}
\bibinfo{author}{Canale, C.}, \bibinfo{author}{Torre, B.},
  \bibinfo{author}{Ricci, D.} \& \bibinfo{author}{Braga, P.~C.}
\newblock \emph{\bibinfo{title}{Recognizing and Avoiding Artifacts in Atomic
  Force Microscopy Imaging}}, \bibinfo{pages}{31--43}
  (\bibinfo{publisher}{Humana Press}, \bibinfo{address}{Totowa, NJ},
  \bibinfo{year}{2011}).

\bibitem{quaroni2018}
\bibinfo{author}{Quaroni, L.}, \bibinfo{author}{Pogoda, K.},
  \bibinfo{author}{Wiltowska-Zuber, J.} \& \bibinfo{author}{Kwiatek, W.~M.}
\newblock \bibinfo{title}{Mid-infrared spectroscopy and microscopy of
  subcellular structures in eukaryotic cells with atomic force microscopy –
  infrared spectroscopy}.
\newblock \emph{\bibinfo{journal}{RSC Adv.}} \textbf{\bibinfo{volume}{8}},
  \bibinfo{pages}{2786--2794} (\bibinfo{year}{2018}).

\bibitem{dazzi2012}
\bibinfo{author}{Dazzi, A.} \emph{et~al.}
\newblock \bibinfo{title}{Afm–ir: Combining atomic force microscopy and
  infrared spectroscopy for nanoscale chemical characterization}.
\newblock \emph{\bibinfo{journal}{Applied Spectroscopy}}
  \textbf{\bibinfo{volume}{66}}, \bibinfo{pages}{1365--1384}
  (\bibinfo{year}{2012}).

\bibitem{Wuebbeler2021}
\bibinfo{author}{Wübbeler, G.}, \bibinfo{author}{Marschall, M.},
  \bibinfo{author}{Rühl, E.}, \bibinfo{author}{Kästner, B.} \&
  \bibinfo{author}{Elster, C.}
\newblock \bibinfo{title}{Compressive nano-{FTIR} chemical mapping}.
\newblock \emph{\bibinfo{journal}{Measurement Science and Technology}}
  \textbf{\bibinfo{volume}{33}}, \bibinfo{pages}{035402}
  (\bibinfo{year}{2021}).

\bibitem{marschall2020}
\bibinfo{author}{Marschall, M.} \emph{et~al.}
\newblock \bibinfo{title}{{Compressed FTIR spectroscopy using low-rank matrix
  reconstruction}}.
\newblock \emph{\bibinfo{journal}{Optics Express}}
  \textbf{\bibinfo{volume}{28}}, \bibinfo{pages}{38762} (\bibinfo{year}{2020}).

\bibitem{Candes2008}
\bibinfo{author}{Candes, E.~J.} \& \bibinfo{author}{Recht, B.}
\newblock \bibinfo{title}{{Exact low-rank matrix completion via convex
  optimization}}.
\newblock In \emph{\bibinfo{booktitle}{2008 46th Annual Allerton Conference on
  Communication, Control, and Computing}}, \bibinfo{pages}{806--812}
  (\bibinfo{publisher}{IEEE}, \bibinfo{year}{2008}).

\bibitem{Davenport2016}
\bibinfo{author}{Davenport, M.~A.} \& \bibinfo{author}{Romberg, J.}
\newblock \bibinfo{title}{{An Overview of Low-Rank Matrix Recovery From
  Incomplete Observations}}.
\newblock \emph{\bibinfo{journal}{IEEE Journal of Selected Topics in Signal
  Processing}} \textbf{\bibinfo{volume}{10}}, \bibinfo{pages}{608--622}
  (\bibinfo{year}{2016}).

\bibitem{WHO2022}
\bibinfo{title}{Leishmaniasis}  (\bibinfo{year}{2022}).
\newblock
  \urlprefix\url{https://www.who.int/en/news-room/fact-sheets/detail/leishmaniasis}.

\bibitem{mathurin2022}
\bibinfo{author}{Mathurin, J.} \emph{et~al.}
\newblock \bibinfo{title}{Photothermal {AFM-IR} spectroscopy and imaging:
  Status, challenges, and trends}.
\newblock \emph{\bibinfo{journal}{Journal of Applied Physics}}
  \textbf{\bibinfo{volume}{131}}, \bibinfo{pages}{010901}
  (\bibinfo{year}{2022}).

\bibitem{Rajes2021}
\bibinfo{author}{Rajes, K.} \emph{et~al.}
\newblock \bibinfo{title}{Oxidation-sensitive core-multishell nanocarriers for
  the controlled delivery of hydrophobic drugs}.
\newblock \emph{\bibinfo{journal}{ACS Biomater. Sci. Eng.}}
  \textbf{\bibinfo{volume}{7}}, \bibinfo{pages}{2485--2495}
  (\bibinfo{year}{2021}).

\bibitem{Metzner2022}
\bibinfo{author}{Metzner, S.} \emph{et~al.}
\newblock \bibinfo{title}{{Assessment of sub-sampling schemes for compressive
  nano-FTIR imaging}} \bibinfo{pages}{1--7} (\bibinfo{year}{2022}).
\newblock \urlprefix\url{http://arxiv.org/abs/2204.05531}.
\newblock \eprint{2204.05531}.

\bibitem{kaestner2018}
\bibinfo{author}{Kästner, B.} \emph{et~al.}
\newblock \bibinfo{title}{Compressed sensing ftir nano-spectroscopy and
  nano-imaging} \textbf{\bibinfo{volume}{26}}, \bibinfo{pages}{18115--18124}
  (\bibinfo{year}{2018}).

\bibitem{chen2020}
\bibinfo{author}{Chen, Y.}, \bibinfo{author}{Huang, T.-Z.},
  \bibinfo{author}{He, W.}, \bibinfo{author}{Yokoya, N.} \&
  \bibinfo{author}{Zhao, X.-L.}
\newblock \bibinfo{title}{Hyperspectral image compressive sensing
  reconstruction using subspace-based nonlocal tensor ring decomposition}.
\newblock \emph{\bibinfo{journal}{IEEE Transactions on Image Processing}}
  \textbf{\bibinfo{volume}{29}}, \bibinfo{pages}{6813--6828}
  (\bibinfo{year}{2020}).

\bibitem{weng2012low}
\bibinfo{author}{Weng, Z.} \& \bibinfo{author}{Wang, X.}
\newblock \bibinfo{title}{Low-rank matrix completion for array signal
  processing}.
\newblock In \emph{\bibinfo{booktitle}{2012 IEEE International Conference on
  Acoustics, Speech and Signal Processing (ICASSP)}},
  \bibinfo{pages}{2697--2700} (\bibinfo{organization}{IEEE},
  \bibinfo{year}{2012}).

\bibitem{peng2014reweighted}
\bibinfo{author}{Peng, Y.}, \bibinfo{author}{Suo, J.}, \bibinfo{author}{Dai,
  Q.} \& \bibinfo{author}{Xu, W.}
\newblock \bibinfo{title}{Reweighted low-rank matrix recovery and its
  application in image restoration}.
\newblock \emph{\bibinfo{journal}{IEEE transactions on cybernetics}}
  \textbf{\bibinfo{volume}{44}}, \bibinfo{pages}{2418--2430}
  (\bibinfo{year}{2014}).

\bibitem{yao2015fast}
\bibinfo{author}{Yao, Q.}, \bibinfo{author}{Kwok, J.~T.} \&
  \bibinfo{author}{Zhong, W.}
\newblock \bibinfo{title}{Fast low-rank matrix learning with nonconvex
  regularization}.
\newblock In \emph{\bibinfo{booktitle}{2015 IEEE International conference on
  data mining}}, \bibinfo{pages}{539--548} (\bibinfo{organization}{IEEE},
  \bibinfo{year}{2015}).

\bibitem{engl1996regularization}
\bibinfo{author}{Engl, H.~W.}, \bibinfo{author}{Hanke, M.} \&
  \bibinfo{author}{Neubauer, A.}
\newblock \emph{\bibinfo{title}{Regularization of inverse problems}}, vol.
  \bibinfo{volume}{375} (\bibinfo{publisher}{Springer Science \& Business
  Media}, \bibinfo{year}{1996}).

\bibitem{Software1}
\bibinfo{author}{Marschall, M.}, \bibinfo{author}{W{\"u}bbeler, G.} \&
  \bibinfo{author}{Elster, C.}
\newblock \bibinfo{title}{Regression--working group 8.42}.
\newblock
  \bibinfo{howpublished}{\url{https://www.ptb.de/cms/nc/en/ptb/fachabteilungen/abt8/fb-84/ag-842/regression-842.html}}
  (\bibinfo{year}{2021}).
\newblock \bibinfo{note}{Accessed: 2021-12-14}.

\bibitem{mahesh2018}
\bibinfo{author}{Mahesh, S.}, \bibinfo{author}{Tang, K.-C.} \&
  \bibinfo{author}{Raj, M.}
\newblock \bibinfo{title}{{Amide Bond Activation of Biological Molecules}}.
\newblock \emph{\bibinfo{journal}{Molecules}} \textbf{\bibinfo{volume}{23}},
  \bibinfo{pages}{2615} (\bibinfo{year}{2018}).

\bibitem{yang2015}
\bibinfo{author}{Yang, H.}, \bibinfo{author}{Yang, S.}, \bibinfo{author}{Kong,
  J.}, \bibinfo{author}{Dong, A.} \& \bibinfo{author}{Yu, S.}
\newblock \bibinfo{title}{{Obtaining information about protein secondary
  structures in aqueous solution using Fourier transform IR spectroscopy}}.
\newblock \emph{\bibinfo{journal}{Nature Protocols}}
  \textbf{\bibinfo{volume}{10}}, \bibinfo{pages}{382--396}
  (\bibinfo{year}{2015}).

\bibitem{parker1975}
\bibinfo{author}{Parker, F.~S.}
\newblock \bibinfo{title}{Biochemical applications of infrared and raman
  spectroscopy}.
\newblock \emph{\bibinfo{journal}{Applied Spectroscopy}}
  \textbf{\bibinfo{volume}{29}}, \bibinfo{pages}{129--147}
  (\bibinfo{year}{1975}).

\bibitem{helm1991}
\bibinfo{author}{Helm, D.}, \bibinfo{author}{Labischinski, H.},
  \bibinfo{author}{Schallehn, G.} \& \bibinfo{author}{Naumann, D.}
\newblock \bibinfo{title}{Classification and identification of bacteria by
  fourier-transform infrared spectroscopy}.
\newblock \emph{\bibinfo{journal}{Microbiology}}
  \textbf{\bibinfo{volume}{137}}, \bibinfo{pages}{69--79}
  (\bibinfo{year}{1991}).

\bibitem{ward1963}
\bibinfo{author}{{Ward Jr.}, J.~H.}
\newblock \bibinfo{title}{Hierarchical grouping to optimize an objective
  function}.
\newblock \emph{\bibinfo{journal}{Journal of the American Statistical
  Association}} \textbf{\bibinfo{volume}{58}}, \bibinfo{pages}{236--244}
  (\bibinfo{year}{1963}).

\bibitem{hornemann2017}
\bibinfo{author}{Hornemann, A.} \emph{et~al.}
\newblock \bibinfo{title}{{A pilot study on fingerprinting Leishmania species
  from the Old World using Fourier transform infrared spectroscopy}}.
\newblock \emph{\bibinfo{journal}{Anal. Bioanal. Chem.}}
  \textbf{\bibinfo{volume}{409}}, \bibinfo{pages}{6907} (\bibinfo{year}{2017}).

\bibitem{centrone2015}
\bibinfo{author}{Centrone, A.}
\newblock \bibinfo{title}{Infrared imaging and spectroscopy beyond the
  diffraction limit}.
\newblock \emph{\bibinfo{journal}{Annual Review of Analytical Chemistry}}
  \textbf{\bibinfo{volume}{8}}, \bibinfo{pages}{101--126}
  (\bibinfo{year}{2015}).

\bibitem{wheeler2013}
\bibinfo{author}{Wheeler, R.~J.}, \bibinfo{author}{Gluenz, E.} \&
  \bibinfo{author}{Gull, K.}
\newblock \bibinfo{title}{The limits on trypanosomatid morphological
  diversity}.
\newblock \emph{\bibinfo{journal}{PloS one}} \textbf{\bibinfo{volume}{8}},
  \bibinfo{pages}{e79581--e79581} (\bibinfo{year}{2013}).

\bibitem{Mathematica}
\bibinfo{author}{{Wolfram Research{,} Inc.}}
\newblock \bibinfo{title}{Mathematica, {V}ersion 11}.
\newblock \urlprefix\url{https://www.wolfram.com/mathematica}.
\newblock \bibinfo{note}{Champaign, IL, 2021}.

\bibitem{levenberg1944}
\bibinfo{author}{Levenberg, K.}
\newblock \bibinfo{title}{A method for the solution of certain non-linear
  problems in least squares}.
\newblock \emph{\bibinfo{journal}{Quarterly of Applied Mathematics}}
  \textbf{\bibinfo{volume}{2}}, \bibinfo{pages}{164--168}
  (\bibinfo{year}{1944}).

\bibitem{marquardt1963}
\bibinfo{author}{Marquardt, D.~W.}
\newblock \bibinfo{title}{An algorithm for least-squares estimation of
  nonlinear parameters}.
\newblock \emph{\bibinfo{journal}{Journal of the Society for Industrial and
  Applied Mathematics}} \textbf{\bibinfo{volume}{11}},
  \bibinfo{pages}{431--441} (\bibinfo{year}{1963}).

\end{thebibliography}

\newpage
\noindent{\bf APPENDIX}
\newline
\newline
{\bf Methods}

\noindent {\textit{AFM-IR nanoscopy on single parasites}}
\newline
For the studies on single \textit{Leishmania brasiliensis} a 450 $\mu$m long gold-coated silicon AFM tip with a nominal radius of 25$\,$nm (Bruker, model: PR-EX-nIR2-10, resonance frequency of 13 $\pm$ 4$\,$kHz, spring constant: 0.07 - 0.4$\,$Nm$^{-1}$) was used. The AFM-IR spectra were acquired within a range of 1306$\,$–$\,$1894$\,$cm$^{-1}$, employing a quantum cascade laser MIRcat-QT\textsuperscript{TM} (DRS Daylight Solutions, USA) equipped with four diode modules. A sweep speed of 100$\,$cm$^{-1}$/s, and a spectral resolution of 4$\,$cm$^{-1}$ were used. The average of the infrared laser power applied to the sample was of 1.5~$\,$mW. The influence of power variation in the IR source at different wavenumbers was reduced by collecting the laser power spectrum and normalizing the AFM-IR amplitude at the specified wavenumbers. The hyperspectral imaging infrared maps were recorded in the resonance enhanced mode with a pixel density of 0.04$\,$$\mu$m$^2$/px$^2$.

In the resonance enhanced mode the repetition rate of the QCL laser is tuned to the continuously monitored AFM cantilever oscillation amplitude within a range of $\pm$ 30$\,$kHz of oscillation resonance. The oscillation resonance during the scan has a slightly higher frequency than the cantilever's free resonance in air and is continuously changing. This is due to the fact that the contact resonance of the cantilever in a snap-in position of the probing tip is varying and these changes are caused by numerous factors such as the contact surface between the probing tip and the investigated sample, the free resonance frequency of the cantilever, or stiffness of the sample. Thus, the oscillation resonance has to be tuned more frequently for complex samples.
\vspace*{1cm}

\noindent {\textit{Cultivation and preparation of Leishmania parasites}}
\noindent 
\newline
Promastigote forms of \textit{Leishmania (V.) brasiliensis} strain (MHOM/PE/94/LC2452cl3) were cultured at 24$^\circ$C ± 1 $^\circ$C in M199 medium (Sigma Aldrich) supplemented with 10\% fetal calf serum (Sigma Aldrich), 1\% L-glutamine, and 0.5\% penicillin/streptomycin (Sigma Aldrich). A neutral pH was ensured by the addition of 0.04M HEPES-NaOH buffer solution (pH 6.9). 
Parasites were collected between the 4\textsuperscript{th} and 5\textsuperscript{th} day in culture. The parasites' density was determined with a Casy counter (OLS OMNI Life Science) and concentration adjusted to 5×10\textsuperscript{6}/mL in order to receive thin films for AFM-IR investigations. After two washing steps with 0.5\% phosphate-buffered saline (PBS) at 1000 g, for 10 min, parasites were resuspended in 300 µL of 0.5\% PBS (or 100 µL for thick drops). A 5 µL drop of the suspension was placed onto a Kevley low-e-slide (Kevley Technologies\textsuperscript{\textcopyright}) and left to air-dry at room temperature. The parasites were microscopically examined with respect to their living or life cycle stages \cite{wheeler2013}. This should ensure that the parasites shall stay under near-to-native condition, when prepared and air-dried onto the substrates. The applied salt concentration was carefully evaluated and compromised with regard to keeping the parasites alive during preparation steps. We would like to stress here that no other additives, i.e. fetal calf serum (FCS) or bovine serum albumin (BSA), were used, as they contain proteins and may render the IR investigations unreliable with regard to protein detection in parasites. 

\vspace*{1cm}

\noindent {\textit{Implementation of Wards algorithm}}
\noindent 
\newline
 The cluster analysis based on Ward's method \cite{ward1963} has been implemented using Mathematica~\cite{Mathematica}. Prior to clustering the data were pre-processed by smoothing using the Savitzky-Golay-Filter with a window size of 10 pixel and using 2\textsuperscript{nd} order polynomials and subsequently taking the 2\textsuperscript{nd} derivative. The hierachical clustering was performed using Ward's minimum variance criterion. The number of clusters was manually limited to three clusters.

\vspace*{1cm}

\noindent {\textit {Nonlinear curve fitting to Voigt Profiles}}
\noindent 
\newline
The applied Voigt profiles constitute a combination of Lorentz and Gauss functions which is taken to approximate the amide I and amide II bands.  
We applied an iterative curve fit procedure to the full and reconstructed spectra implementing the Levenberg-Marquardt algorithm \cite{levenberg1944,marquardt1963}, and we paid attention to the fit converging until $\chi$\textsuperscript{2} does not change anymore (relative tolerance lying at 1$\times$10\textsuperscript{-9}). In addition, the same boundaries of the amide peak flanks were considered for the respective fitting case.




\end{document}